\documentclass[aps,
twocolumn,
amsmath,amssymb,
prc, 
floatfix,
preprintnumbers,
nofootinbib,
superscriptaddress
]{revtex4-2}

\usepackage{graphicx}
\usepackage{dcolumn}
\usepackage{bm}
\usepackage{color}
\usepackage[dvipsnames]{xcolor}
\usepackage[colorlinks=true,allcolors=blue]{hyperref}
\usepackage{braket}
\usepackage{amsmath,amssymb,amsthm}
\usepackage[normalem]{ulem}

\newcommand{\xvec}[1]{\boldsymbol{#1}}

\newcolumntype{Y}{>{\centering\arraybackslash}X}

\renewcommand{\paragraph}[1]{\vspace{3pt}\noindent \textbf{#1} ---}

\begin{document}

\preprint{LA-UR-26-28248}
\preprint{INT-PUB-26-031}

\title{Parametric Neural Quantum States for Nuclear Physics}

\author{Yukari Yamauchi}
\email{yyamauchi@lanl.gov}
\affiliation{Theoretical Division, Los Alamos National Laboratory, Los Alamos, NM 87545, USA}

\author{Ryan Curry}
\email{rmcurry@lanl.gov}
\affiliation{Theoretical Division, Los Alamos National Laboratory, Los Alamos, NM 87545, USA}

\author{Garrett B. King}
\email{kinggar@uw.edu}
\affiliation{Institute for Nuclear Theory, University of Washington, Seattle, WA 98195, USA}
\affiliation{Theoretical Division, Los Alamos National Laboratory, Los Alamos, NM 87545, USA}

\author{Scott Lawrence}
\email{srlawrence@lanl.gov}
\affiliation{Theoretical Division, Los Alamos National Laboratory, Los Alamos, NM 87545, USA}

\author{Rahul Somasundaram}
\email{rsomasundaram@lanl.gov}
\affiliation{Theoretical Division, Los Alamos National Laboratory, Los Alamos, NM 87545, USA}

\author{Ingo Tews}
\email{itews@lanl.gov}
\affiliation{Theoretical Division, Los Alamos National Laboratory, Los Alamos, NM 87545, USA}

\date{\today}

\begin{abstract}
We introduce a new approach, based on neural quantum states (NQSs), to rapidly compute nuclear observables when couplings in nuclear Hamiltonians are varied.
A single interaction-dependent NQS, trained across a range of couplings in the Hamiltonian, provides high-fidelity wavefunctions for that continuous range of interaction parameters.  
With access to the wavefunction for each set of couplings, any static observables can be computed efficiently without retraining the NQS.
We apply this framework to two-body nuclear scattering and the deuteron ground state with local interactions derived from chiral effective field theory up to third order.
\end{abstract}

\maketitle

\paragraph{Introduction}
Chiral effective field theory (EFT) provides a systematic theory of nuclear forces: an order-by-order expansion scheme to derive nuclear interactions but with an undetermined set of low-energy couplings (LECs)~\cite{Epelbaum:2008ga,Machleidt:2011zz}. 
Determining these LECs requires high-accuracy first-principles calculations of nuclear systems, which can be compared with experimental data to constrain the LEC values.
Commonly, such fitting procedures use two-nucleon scattering data~\cite{Entem:2003ft,Epelbaum:2004fk, Epelbaum:2014efa, Gezerlis:2013ipa, Piarulli:2014bda,Epelbaum:2014sza,Entem:2017gor, Reinert:2017usi, Somasundaram:2023sup} and the properties of light atomic nuclei~\cite{Nogga:2001cz,Navratil:2007we, Hebeler:2010xb, Lynn:2015jua, Baroni:2016xll, LENPIC:2018ewt, Curry:2025pna} ($A\leq 5$), for which precision simulations are feasible. 
Alternatively, nuclear interactions have also been adjusted to properties of medium-mass nuclei~\cite{Ekstrom:2015rta, Hu:2021trw, Arthuis:2024mnl, Hu:2025cjl} or even of neutron stars~\cite{Somasundaram:2024ykk,Armstrong:2026aap}, as these systems provide complementary information. 

These fitting strategies require highly accurate and precise many-body methods to connect LECs to data.
A robust calibration of LECs also requires that physical observables be computed for a large number of sets of couplings.
This is expensive: even when a high-precision computation is tractable for a single set of LECs, it may be prohibitively expensive to perform a fit based on many evaluations.
This challenge motivates the use of surrogate models~\cite{Frame:2017fah,Konig:2020,Bonilla:2022,Giuliani:2023,Melendez:2022kid,Drischler:2022ipa,Cook:2024toj,Belley:2025nkn,Heihoff:2026ycq} that speed up quantum many-body calculations at the cost of small emulation errors. 
Such emulators have enabled sensitivity studies of Hamiltonian operators~\cite{Ekstrom:2019lss} as well as Bayesian fits of LECs to different many-body observables~\cite{Curry:2025pna,Armstrong:2026aap}.
A drawback of data-driven emulators, such as parametric matrix models~\cite{Cook:2024toj}, is that they must be trained separately for each observable, even for the same underlying system. 

In this Letter, we present a new high-fidelity approach, based on neural quantum states (NQSs), to rapidly compute nuclear observables for many sets of LECs.
NQS methods, first proposed in Ref.~\cite{Carleo:2016svm}, use neural networks to parametrize a wavefunction. 
They have been used to study ground-state properties of atomic nuclei up to medium mass with nuclear forces inspired by pionless EFT at leading order (LO)~\cite{Gnech:2023prs,Fore:2026spn}, nuclei up to $A=3$ for more realistic chiral EFT interactions~\cite{Wen:2025mlq}, and nuclear scattering~\cite{Lawrence:2026dnb} with a phenomenological interaction. 

These NQS approaches were constructed for fixed nuclear Hamiltonians.
The NQSs developed in this work, instead, are built to include LECs of nuclear interactions as input parameters, acting directly as fast high-fidelity solvers as couplings are varied.
Because NQSs output the wavefunction itself, rather than observables alone, they can be used to extract expectation values of arbitrary operators, often with well-bounded uncertainties.
We train NQSs for the deuteron and two-nucleon scattering, using local chiral EFT interactions up to next-to-next-to-leading order (N$^2$LO)~\cite{Somasundaram:2023sup}.
All demonstrations presented in this project make use of the Python packages \texttt{jax}~\cite{deepmind2020jax} and \texttt{equinox}~\cite{kidger2021equinox}.
Codes developed for the project and data necessary for reproducing the results are available at Ref.~\cite{chineura}. 

\paragraph{Neural Quantum States}
NQSs parameterize wavefunctions using neural networks, exploiting their flexibility to provide accurate high-dimensional representations of quantum states~\cite{pfau2020ferminet}. 
NQS methods have been applied to nuclear ground states~\cite{Gnech:2023prs,Wen:2025mlq,Fore:2026spn}, various response functions~\cite{Parnes:2025seu}, and scattering~\cite{Lawrence:2026wrf,Lawrence:2026dnb}. 
When applied to nuclear systems, the NQSs usually take as input the positions of $A$ nucleons and output the value of the wavefunction for the given configuration. 
The dependence of the wavefunction on internal degrees of freedom, such as spin and isospin, can be incorporated in either the input or the output. 
In this work, the NQS outputs $2^A$ (with $A=2$) complex numbers to account for the spin dependence.
The isospin degree of freedom is explicitly implemented by building the Hilbert space as a tensor product of those of distinguishable protons and neutrons.

The NQS is then trained according to an appropriate loss function, so it represents a bound state or a scattering state that approximately satisfies the time-independent Schr{\"o}dinger equation. 
Denoting the NQS as $\Psi_\alpha(\xvec{x})$ with positions $\xvec{x} \in \mathbb{R}^{3A}$ and neural-network parameters $\alpha$, we use the energy expectation value as our loss function for ground-state calculations:
\begin{equation}\label{eq:deuteron-loss-e}
    L_g\left[\Psi_\alpha\right]  = \frac{\int d\xvec{x} ~ \Psi_\alpha^{\dagger}(\xvec{x})\hat H \Psi_\alpha(\xvec{x})}{\int d\xvec{x} ~ \Psi_\alpha^{\dagger}(\xvec{x}) \Psi_\alpha(\xvec{x})} \,\text.
\end{equation}
For scattering states, the goal is not to minimize the energy but instead to find an eigenstate of known energy with particular asymptotic behavior. 
In this case, we use as our loss function the $L_1$ norm of the violation of the time-independent Schr\"odinger equation, defined as
\begin{equation}
    L_s\left[\Psi_\alpha\right]  = \int d\xvec{x}~ \left| (E - \hat H) \Psi_\alpha(\xvec{x}) \right|\,.
    \label{eq:loss_scatt}   
\end{equation}
This choice was proven to provide a stable variational principle in Ref.~\cite{Lawrence:2026wrf}, in contrast to traditional approaches such as the Kohn variational principle~\cite{Kohn:1948col}. 

Each NQS calculation of either a nuclear ground state or a scattering state costs considerable computational resources: for example, a single deuteron or neutron-proton scattering calculation in Ref.~\cite{Lawrence:2026dnb} took $\sim 1$ hour on an M2 processor.
For applications where many sets of LECs, and hence, many different wavefunctions, need to be evaluated, such NQS computations may become infeasible.
This hinders, e.g., applications of NQS to Bayesian calibration of nuclear interactions.
To solve this problem, here we develop \textit{LEC-dependent NQSs}. 
These NQSs take as input not only the positions of nucleons but also values of LECs and other parameters that define nuclear interactions and properties of the state of interest, e.g., the relative momentum of two incoming nucleons for scattering states. 
Consequently, a trained NQS will represent a system of interest for a range of reasonable LECs obtained from state-of-the-art (e.g., Bayesian) calibrations. 
By repeatedly calling the trained NQS, we are able to compute nuclear observables for any set of LECs within the trained region of parameter space, thus circumventing the need for training a large number of individual NQSs. 
In this work, we demonstrate this LEC-dependent NQS approach for two systems: the deuteron ground state and the neutron-proton elastic scattering state. 

\begin{figure}[t]
    \centering
    \includegraphics[width=0.99\linewidth]{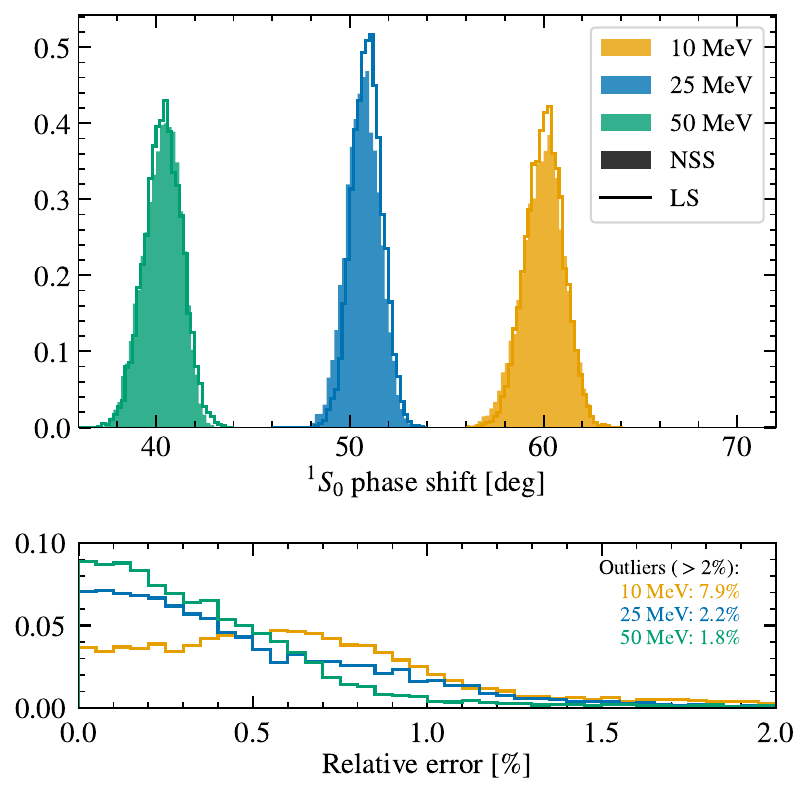}
    \caption{$^1S_0$ phase shifts at lab energies of \mbox{$E=10,25,50$~MeV} computed using the neural scattering state (NSS) and compared with results obtained by solving the Lippmann-Schwinger (LS) equation for 10,000 LEC samples from the N$^2$LO posterior of Ref.~\cite{Somasundaram:2023sup}. 
    In the lower panel, we show the distribution of relative errors for the NSS, and state the fraction of samples with relative errors larger than 2\%.
    }
    \label{fig:npscatter}
\end{figure}

\paragraph{Nuclear interactions}
Our LEC-dependent NQSs can be trained for any family of nuclear Hamiltonians. 
In this work, we employ the maximally-local chiral EFT two-nucleon interactions of Ref.~\cite{Somasundaram:2023sup} up to N$^2$LO. 
These Hamiltonians consist of longer-range pion-exchange contributions as well as short-range contact interactions. 
In this work, we fix the couplings in the pion-exchange terms to the same values chosen in Ref.~\cite{Somasundaram:2023sup} and vary only the short-range LECs.
At LO, there are two such contacts as well as the well-known one-pion-exchange interaction. 
At NLO and N$^2$LO, there are seven additional LECs in the short-range potential, so a total of nine must be determined by fitting. 
There are also additional two-pion exchange contributions appearing at these orders with radial functions given in Ref.~\cite{Gezerlis:2014zia}.
In Ref.~\cite{Somasundaram:2023sup}, the contact LECs were fit to phase shifts using a Bayesian fitting protocol which resulted in LEC posteriors that include estimates for the theoretical uncertainties of the interactions. 
Here, we employ these interactions at the cutoff $R_0=0.6$~fm and train the NQS on LEC ranges supported by the posteriors from Ref.~\cite{Somasundaram:2023sup}.

The training procedure for the parameter-dependent NQSs requires us to have a sampler for the LEC space that is fast and unbiased. 
Moreover, with our present approach, it is essential that we train the NQS on a physically sensible domain of the LEC space. 
For example, when training the NQS for the deuteron, it is necessary that a bound state exists for all LEC sets in the training domain to ensure the stability of the training process.
To impose such constraints on the LEC sampler, here, we randomly drew $272,000$ LEC sets from the posterior of Ref.~\cite{Somasundaram:2023sup} in advance, and the LEC sampler randomly draws from this list. 
The LEC ranges for training are given in Table~\ref{table:lec} but we stress that these do not reflect the correlations among the LECs~\cite{Dezdarani:2025aws}.
Such a sampling procedure may introduce unpredictable biases in the training, depending on the quality of the pre-collected samples. 
For a more reliable training of NQSs in the LEC space, one may consider using unbiased sampling methods, such as normalizing flows~\cite{Yamauchi:2023xrz}, at each training step.  
We will explore such approaches in future work.

\paragraph{Results for neutron-proton scattering}
We construct neural scattering states (NSSs), a specific version of the NQS, for neutron-proton scattering with total spin~0.
The NSS takes as input the positions of the neutron and proton $\xvec{x_n}$ and $\xvec{x_p} (\in \mathbb{R}^3)$, the interaction parameters $\vec c$, and the scattering momentum $k$ in the center-of-mass frame. 
Following Ref.~\cite{Lawrence:2026dnb}, the scattering state can be decomposed into three terms:
\begin{equation}
    \Psi^{(np)}_{\alpha}(\xvec{x_n}, \xvec{x_p}, \{\vec c,k\}) = \psi_{\mathrm{in}} + \psi_{\mathrm{bulk}} + \psi_{\mathrm{out}} \,. \label{eq:ansatz}
\end{equation}
The incoming state $\psi_{\mathrm{in}}$ is fixed throughout the training, and uniquely identifies the asymptotic behavior of the scattering state~\cite{Lawrence:2026wrf}. 
It is given by
\begin{equation}
    \psi_{\mathrm{in}} = e^{ik(z_p-z_n)} \,u_n \otimes u_p \,,
\end{equation}
where the incoming spins of neutron and proton are denoted as $u_n$ and $u_p$, respectively, and are fixed such that the state has total spin 0.
We do not include the incoming spins as input parameters of the NSS in this work. 

The outgoing state $\psi_{\mathrm{out}}$ encodes the asymptotic behavior of the scattering state, from which one extracts scattering amplitudes:
\begin{equation}\label{eq:np_far}
    \psi_{\mathrm{out}} = \frac{e^{ikr}}{r} \left[ f(\Omega, \{\vec c,k\}) + \frac{\Delta_{\Omega}f}{2ikr} \right] W_{np,\mathrm{out}}(r) \,\text. 
\end{equation}
Here, the relative coordinate of the proton and neutron, $\xvec{x_n}-\xvec{x_p}$,  is expressed in spherical coordinates (distance $r$ and angles $\Omega$) and $\Delta_{\Omega}$ denotes the Laplace-Beltrami operator on the unit sphere.
To keep $L_s$ of Eq.~\eqref{eq:loss_scatt} finite, it is necessary to include terms up to first order in the expansion of the asymptotic state in $1/r$.
In this work, we parametrize the scattering amplitude $f$ with spherical harmonics up to $\ell=2$, as in Ref.~\cite{Lawrence:2026dnb}:
\begin{equation}
    f(\Omega, \{\vec c,k\}) = \sum_{\ell=0,m_{\ell}}^{\ell=2} f_{\ell m_{\ell}}(\{\vec c,k\}) ~ Y_\ell^{m_{\ell}}(\Omega) \;\text.
\end{equation}
For all three orbital angular momenta, we construct neural networks $f_{\ell m_{\ell}}$ that take as input the interaction and scattering parameters, $\{\vec c,k\}$, and output $4\cdot(2l+1) = 4, 12$, or $20$ complex numbers for $\ell=0,1,2$, respectively, to obtain the dependence on the spin and orbital angular momentum $L_z$.
The window function $W_{np,\mathrm{out}}$ in Eq.~(\ref{eq:np_far}) enforces that $\psi_{\mathrm{out}}$ goes smoothly to zero as $r\to 0$. 

The interaction of the nucleons during the scattering process is captured by $\psi_{\mathrm{bulk}}$, which takes as input the array consisting of relative coordinate $\xvec{x_n}-\xvec{x_p}$ and $\{\vec c,k\}$, and outputs the real and imaginary parts of a four-component complex-valued vector in spin space.
For further details of the neural network architecture for $\psi_{\mathrm{out}}$ and $\psi_{\mathrm{bulk}}$, see the supplemental material. 

We train the NQS ansatz of Eq.~\eqref{eq:ansatz} with the loss function defined in Eq.~\eqref{eq:loss_scatt}, which integrates the $L_1$ violation of the Schr{\"o}dinger equation in both coordinate and parameter spaces:
\begin{equation}\label{eq:Ls}
    L_s\left[\Psi_{\alpha}\right]  = \int dk\,d\vec c\,d\xvec{x} \left| (E - \hat H) \Psi_{\alpha}(\xvec{x},\{\vec c,k\}) \right|\,.
\end{equation}
At each training step, we approximate the loss with $2^{14}$ samples in $k, \vec c$, and $\xvec{x}$ space. 
For the parameter space, $k$ is sampled uniformly in the range $[0.245, 0.814]$~fm$^{-1}$, which corresponds to lab energies in the range of $[5, 55]$~MeV. 
The LEC samples are drawn randomly from the $272,000$ sets as discussed previously. 
For the coordinate-space integral, we draw samples from an exponential distribution 
\begin{equation}\label{eq:sampler}
    P(r,\Omega) = r^{-2} e^{-r/R}
\end{equation}
with $R=2.0$ fm, similar to Ref.~\cite{Lawrence:2026dnb};  
see the supplemental material 
for further details of the training. 

\begin{table}[t]
\begin{center}
\caption{Range of LEC values used to train the NQS for the deuteron and the NSS for $np$ scattering for different operators at N$^2$LO with a cutoff of $R_0=0.6$ fm.
Note that these ranges do not reflect correlations among the LECs.}
\label{table:lec}
\begin{tabular*}{\columnwidth}{@{\extracolsep{\fill}}r|c|c|c}
 \hline\hline
 Operator & LEC  & Min & Max  \\
 \hline\hline
 $1$ & $C_S$ [fm$^2$] & 4.70 & 25.5  \\ 
 \hline
 $\bm{\sigma}_1 \cdot \bm{\sigma}_2$ & $C_T$ [fm$^2$] & -0.13 &  6.80  \\
 \hline
 $q^2$ & $C_1$ [fm$^4$]& -0.37 & 0.34   \\
 \hline
 $q^2 \bm{\tau}_1 \cdot \bm{\tau}_2 $ & $C_2$ [fm$^4$] & -0.25 & 0.59  \\
 \hline
 $q^2 \bm{\sigma}_1 \cdot \bm{\sigma}_2 $ & $C_3$ [fm$^4$]& -0.23 & 0.14  \\
 \hline
 $q^2 \bm{\sigma}_1 \cdot \bm{\sigma}_2 \bm{\tau}_1 \cdot \bm{\tau}_2 $ & $C_4$ [fm$^4$]& -0.06 & 0.52  \\
 \hline
 $i/2 (\bm{\sigma}_1 + \bm{\sigma}_2) \cdot \bm{q} \times \bm{k} $ & $C_5$ [fm$^4$]&  -5.00 & -1.74  \\
 \hline
 $(\bm{\sigma}_1 \cdot \bm{q}) (\bm{\sigma}_2 \cdot \bm{q}) $ & $C_6$ [fm$^4$]& -0.15 & 0.78  \\
 \hline
  $(\bm{\sigma}_1 \cdot \bm{q}) (\bm{\sigma}_2 \cdot \bm{q}) \bm{\tau}_1 \cdot \bm{\tau}_2 $ & $C_7$ [fm$^4$]& -1.00 & -0.03 \\
\end{tabular*}
\end{center}
\end{table}

The trained NSS represents the scattering states for a range of LECs in the nuclear Hamiltonian as well as $k$.  
From the NSS, all quantities associated with the $S$ matrix can be straightforwardly extracted, including phase shifts, differential cross sections, and total cross section.
Importantly, systematic uncertainties in these observables are quantifiable, as discussed in Ref.~\cite{Lawrence:2026wrf}, due to the availability of the entire scattering wavefunction.
For example, systematic uncertainties of phase shifts can be bounded from above via Theorem 4 of Ref.~\cite{Lawrence:2026wrf}, or approximated cheaply by observing the violation of unitarity of the S-matrix, as demonstrated in Section 3 of Ref.~\cite{Lawrence:2026dnb}. 
In Fig.~\ref{fig:npscatter}, we show the $^1S_0$ phase shift for lab energies $E_{\mathrm{lab}} = 10, 25$, and $50$~MeV. 
The error is computed by comparing with the exact solutions from the Lippmann-Schwinger (LS) equation~\cite{Lippmann:1950zz}. 
The relative systematic errors of the NQS are $\lesssim 1\%$, and the distributions of phase shifts quantitatively agree with those of LS solutions. 
The training of the NQS used in Fig.~\ref{fig:npscatter} took $9$ hours on an M2 processor, representing a substantial speedup over what would be required to train 10,000 individual NSSs.

\begin{figure}[t]
    \centering
    \includegraphics[width=0.945\linewidth]{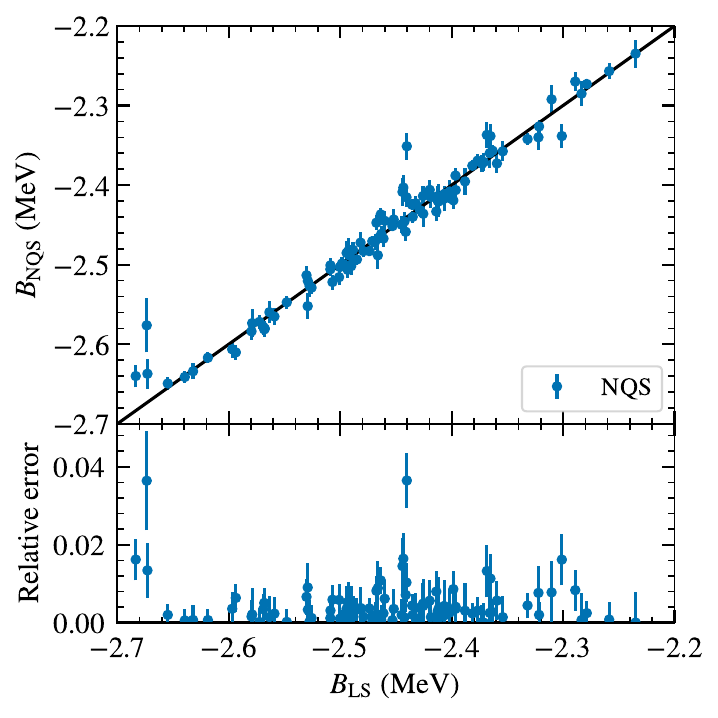}
    \caption{Deuteron ground-state energies predicted by the NQS for 100 parameter sets at N$^2$LO compared to results obtained by solving the Lippmann-Schwinger (LS) equation. 
    }
    \label{fig:deuteron-binding}
\end{figure}

\paragraph{Results for the deuteron}
The NQS for the deuteron $\Psi_{\alpha}^{(d)}(\xvec{x_n}, \xvec{x_p}, \vec c)$ takes as input the positions of the neutron $\xvec{x_n}$ and the proton $\xvec{x_p}$, as well as the LECs, which we again denote collectively as $\vec c$. 
Focusing on the deuteron state with total angular momentum $J_z=+1$, the wavefunction can be parametrized by four terms: 
\begin{align}\label{eq:psid}
    \Psi_{\alpha}^{(d)}(\xvec{x_n}, \xvec{x_p}, \vec c) &=  W_d(r,\vec c)  \Bigl[ f_{0,0}(r,\vec c) ~  Y_0^0(\Omega) ~ |1,1\rangle \nonumber\\
     + \sum_{m_\ell=0,1,2} & f_{2,m_\ell}(r,\vec c) ~  Y_2^{m_\ell}(\Omega) ~ |1,1-m_\ell\rangle \Bigr]\text.
\end{align}
Here, the total spin configurations of the neutron and proton are denoted by $|s, m_s\rangle$.
We parametrize the four complex-valued functions $f_{\ell,m_\ell}(r,\vec c)$ by a single neural network.
This neural network takes as input $[r,r^2,\vec c]$ and outputs the real and imaginary parts of 4 complex numbers, interpreted as the evaluations of $f_{0,0}$ and $f_{2,m_\ell}$. 
The window function $W_d(r)$ characterizes the exponential decay in the tail of the deuteron wavefunction, and is parametrized by another neural network; see the supplemental material
for further details of the neural-network architecture. 

We train the deuteron NQS using the loss function of Eq.~(\ref{eq:deuteron-loss-e}), i.e., we minimize the expectation value of the energy. 
As expected from the variational principle, this yields a wavefunction with energy no less than, but close to, the true ground-state energy.
In principle, one can measure the variance of the Hamiltonian to derive systematic error bounds on the binding energy; however, we found these bounds too loose to be of practical use, even after performing further gradient descent to reduce the variance itself.
At each training step, the loss is estimated with finite samples taken from the LEC posterior~\cite{Somasundaram:2023sup} and the coordinate sampler in Eq.~(\ref{eq:sampler}) with $R=3.33\,\mathrm{fm}$, which we found to give low variance in the estimate of the loss.
For further details on the training, see the supplemental material.

\begin{figure}[t]
    \centering
    \includegraphics[width=\columnwidth]{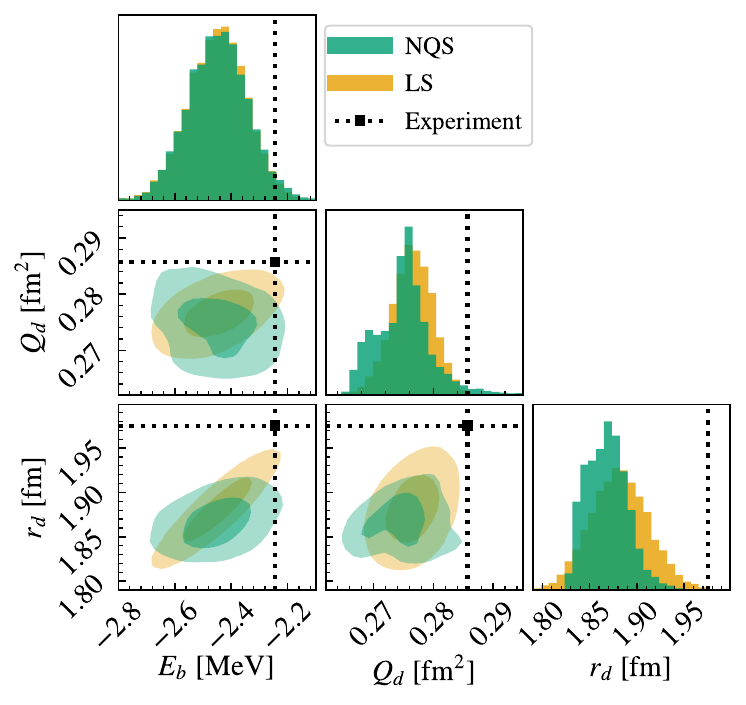}
    \caption{Posteriors for the deuteron ground-state energy $E_b$, the quadrupole moment $Q_d$, and the point-nucleon radius $r_d$ computed for 10,000 parameter sets at N$^2$LO via the single NQS described in the text. 
    The experimental values for the point-nucleon radius and quadrupole moment are taken from Ref.~\cite{Friar:1997js}. 
    The shaded regions in the two-dimensional posteriors capture 50\% and 90\% of the total probability mass. }
    \label{fig3}
\end{figure}

We show the deuteron ground-state energy $E_b$ in Fig.~\ref{fig:deuteron-binding}, where we compare NQS and LS solutions for 100 parameter sets.
Typical relative errors are of the order of 1\%.
The trained NQS captures the deuteron wavefunction for varying LECs; thus, any physical information can be extracted from the NQS by computing expectation values of the corresponding operators. 
This dramatically reduces the computational cost of fitting the LECs. 
To demonstrate this advantage, in addition to the deuteron ground-state energy $E_b$, we have computed its quadrupole moment, defined as $Q_d = 3 \langle\hat z^2\rangle - \langle \hat r^2\rangle$, as well as the point-nucleon radius, defined as $r_d=\sqrt{\langle \hat r^2\rangle}$.  
After the training of the NQS, each evaluation for a single LEC set took approximately $20\,\mathrm{ms}$ on an M2 chip. 
Results are shown in Fig.~\ref{fig3}.

The posterior distribution of LECs was determined exclusively by fitting to nucleon-nucleon scattering data, and we find that the deuteron is systematically overbound, with a point-nucleon radius too small by around $5\%$, consistent with the results of Ref.~\cite{Somasundaram:2023sup}. 
The relative errors of NQS and LS solutions on $Q_d$ and $r_d$ are of comparable size to the relative errors on $E_b$. 
However, because the posterior distributions of these observables are substantially tighter than that of the binding energy, the relative errors on the distributions of $r_d$ and $Q_d$ are larger than the relative error on the distribution of $E_b$.
We anticipate that a trained NQS of this sort can be used to rapidly tune LECs on additional observables, without requiring additional high-cost calculations once training is complete.

\paragraph{Summary and Outlook}
The parameter-dependent NQSs, developed in this work, can serve as ``high-fidelity emulators'' by providing wavefunctions and associated observables for a system of interest over a range of LECs at low computational cost. 
We have demonstrated the capabilities of such NQSs in two systems: neutron-proton scattering and the deuteron.
Observables in these two systems are important quantities commonly used for fitting nuclear interactions. 
Going forward, the parameter-dependent NQSs will be extended to larger systems, in particular to ground states of various light nuclei. 
Such NQSs will provide nuclear observables for a range of couplings, providing a robust computational framework for future Bayesian calibrations of nuclear interactions. 

\paragraph{Acknowledgments}
This document has been approved for unlimited release, and was assigned LA-UR-26-28248. 
This work was supported by the U.S.~Department of Energy through Los Alamos National Laboratory (LANL). 
LANL is operated by Triad National Security, LLC, for the National Nuclear Security Administration of the U.S.~Department of Energy (Contract No.~89233218CNA000001).
Y.Y.~was supported by a Darleane C.~Hoffman fellowship from the Laboratory Directed Research and Development (LDRD) program of LANL under project number 20251144PRD1.
R.C.~and I.T.~were supported by the U.S.~Department of Energy, Office of Science, Office of Nuclear Physics program under Award Number DE-SCL0000015.
G.B.K.~was supported by the LDRD program of LANL under project 20240742PRD1 and acknowledges support from the University of Washington Dehmelt Fellowship in Physics.
S.L.~was supported by a Richard P.~Feynman fellowship from the LDRD program of LANL under project number 20230778PRD1. 
S.L., R.S., and I.T.~were supported by the LDRD program of LANL under project number 20260260ER.
This work benefited from discussions at the Institute for Nuclear Theory at the University of Washington during the program INT 26-1: ``Nuclear Hamiltonians for Advancing Nuclear Physics and Beyond''.
This research was supported in part by the INT's U.S. Department of Energy grant No. DE-FG02-00ER41132.

\vspace{-5pt}
\bibliography{refs}


\clearpage

\section*{Supplemental Material}
\appendix
\label{app:nn}

This appendix provides details on the neural-network architectures defining the NQSs described in the main text, as well as on their training procedures. 

\subsection{Neutron-proton scattering}
The NSS for neutron-proton scattering consists of three terms defined in Eq.~(\ref{eq:ansatz}). 
Among those, the asymptotic state $\psi_{\mathrm{out}}$ and the ``interacting'' part $\psi_{\mathrm{bulk}}$ are parametrized by neural networks. 
A total of six simple multi-layer perceptrons (MLPs) are used to parametrize these two pieces.
All of these networks, as detailed below, take as a part of their input the interaction parameters $\vec c$, whose absolute values 
can be as large as $\sim 10^3$ in our unit system where the speed of light $c=197.3$ (MeV fm). 
To avoid numerical instability during the training process, the couplings are normalized and shifted so that the distribution of the training samples ($272,000$ samples taken from the posterior in Ref.~\cite{Somasundaram:2023sup}) has a mean of zero and a standard deviation of one in each dimension.
For the other inputs ($k$ and the coordinates of nucleons), no such processing was necessary.

For $\psi_{\mathrm{out}}$, we train three neural networks that parametrize $f_{\ell m_{\ell}}$ (with $\ell=0,1,2$) defined in Eq.~(\ref{eq:np_far}). 
The neural networks take as input the interaction and scattering parameters, $\{\vec c,k\}$, and each output 8, 24, or 40 real numbers. 
The output then is folded into $4\cdot(2l+1) = 4, 12$, or $20$ complex numbers for $\ell=0,1,2$, respectively, to accommodate the spin degree of freedom and the projection quantum number $m_\ell$ of the orbital angular momentum.
All three neural networks are MLPs with depth 2 and width 100. 
The ``swish'' activation function~\cite{ramachandran2017searchingactivationfunctions} is used between linear transformations, and no final activation function was used after the final linear transformation. 

The window function $W_{np,\mathrm{out}}(r)$ in Eq.~(\ref{eq:np_far}),
\begin{equation}
    W_{np,\mathrm{out}}(r,\{k, \vec c\}) = e^{-r_{\mathrm{far}}(r,\{k, \vec c\})^2/r^2}\,,
\end{equation}
enforces that $\psi_{\mathrm{out}}$ goes smoothly to zero as $r\to 0$. 
The neural network for $r_{\mathrm{far}}$ takes as input the interaction and scattering parameters, $\{\vec c,k\}$, and outputs one real number. 
We parametrize $r_{\mathrm{far}}$ with another MLP with depth 1 and width 100. 
The swish activation function is used between the two linear transformations, and the hyperbolic tangent function is used as the final activation function after the final linear transformation, to enforce the output to fall in the range $[-1,1]$. 
After adding 1.5~fm to the output of the MLP, the parameter $r_{\mathrm{far}}$ is guaranteed to be in the range $[0.5,2.5]$~fm, which we find to be a good choice for the neutron-proton scattering-state ansatz. 

The interactions of the nucleons during the scattering process are captured by $\psi_{\mathrm{bulk}}$. 
This neural network that parametrizes $\psi_{\mathrm{bulk}}$ takes as input the array consisting of relative coordinate $\xvec{x_n}-\xvec{x_p}$ and $\{\vec c,k\}$, and outputs the real and imaginary parts of a four-component complex-valued vector in spin space.
The neural network is an MLP with depth 3 and width 300. 
The swish activation function is applied between the linear transformations, and no final activation function is applied. 
The output of this MLP is multiplied by the window function $\frac{1}{1+(r/r_{\mathrm{near}})^4}$ to ensure that $\psi_{\mathrm{bulk}}$ decays sufficiently fast with $r$. 
For the parametrization of $r_{\mathrm{near}}$, we use the same neural network architecture as with $r_{\mathrm{far}}$.

We trained the NQS for $10^4$ steps via the Adam optimizer~\cite{kingma2017adammethodstochasticoptimization} with the learning rate fixed at $10^{-3}$ for the first 5000 steps and then decreased to $10^{-5}$ exponentially during the second half of the training.
At each training step, the loss in Eq.~(\ref{eq:Ls}) was estimated via $2^{14}$ samples taken from the $k, \vec c$, and $x$ spaces.
The LEC samples are taken from the pre-collected sets as discussed previously, and a 10\%-Gaussian noise was added to each parameter at each training step. 
Without this addition of random shifts to the LEC samples, the NQS failed to capture the LEC-dependence of the $^1S_0$ phase shift precisely, but instead captured only the central values.

\subsection{Deuteron}
The deuteron wavefunction in Eq.~(\ref{eq:psid}) is parametrized by two MLPs: one for the complex-valued functions $f_{\ell,m_{\ell}}$ and the other one for the window function $W_d(r)$. 
The LECs that enter these MLPs are normalized and shifted, as was done in the neutron-proton scattering case. 
A total of four complex-valued functions $f_{\ell,m_{\ell}}$ are parametrized all together via one MLP with depth 3 and width 400. 
The inverse tangent function was applied as an activation function between the linear transformations, and no final activation function was used. 
The neural network takes as input $[r,r^2,\vec c]$ and outputs 8 real numbers, which get folded into 4 complex numbers that are assigned to $f_{0,0}$ and $f_{2,m_{\ell}}$. 
The window function $W_d(r)$ characterizes the exponential decay in the tail of the deuteron wavefunction,
\begin{equation}
    W_d(r) = \frac{1}{1 + \frac{r}{r_0}} e^{-r \left[0.232\,\mathrm{fm}^{-1} + 10^{-2}  d_d(\{\vec c,k\}) \right]}\,\text.
\end{equation}
We take $r_0=1\,\mathrm{fm}$.
When $d_d=0$, this gives a decay rate appropriate to a deuteron bound by $\sim 2.22\,\mathrm{MeV}$.
The decay-rate correction $d_d$ is parametrized by the second MLP, which has depth 1 and width 45. 
The inverse tangent function is applied after all linear transformations as activation function. 
With the final activation and multiplication of the output by $10^{-2}$, the decay correction rate is constrained to be in the range $[-\pi/200,\pi/200]$. 
While the range of the decay correction is small compared to the central value 0.232~fm$^{-1}$, we find that this freedom improves the training of the deuteron NQS. 

In order to obtain a deuteron NQS which is valid for the entire posterior distribution of LECs, including the tails of the posterior, we follow an adiabatic training procedure. 
Define $\vec c_0$ to be the LECs of maximal likelihood as given in Table IV of Ref.~\cite{Somasundaram:2023sup}. 
For each sample $\vec c_i$ from the posterior, we define a trajectory
\begin{equation}
    \vec c_i(t) = \vec c_0 + t \left(\vec c_i - \vec c_0\right)\text.
\end{equation}
We train for $12$ different values of $t$: $t = \frac{n}{10}$ for integers $1 \le n \le 12$. 
At each value we perform $1000$ steps of the Adam stochastic optimizer~\cite{kingma2017adammethodstochasticoptimization}. 
In the $n=1$ case, we use an exponentially decaying learning rate from $10^{-3}$ to $10^{-4}$, and the learning rate is held at $10^{-4}$ for the remaining $11$ epochs. 
All these trainings happen with the loss given by Eq.~(\ref{eq:deuteron-loss-e}), estimated with $2^{10}$ samples per step. 
Finally, we perform a lengthy refinement training, with scaling $t = 1.1$, with the same loss function estimated with $2^{14}$ samples.
This training takes $15000$ steps, with an exponentially decaying learning rate from $10^{-4}$ to $10^{-7}$. 
This full training procedure is found to take $\sim 1$ hour running on an NVIDIA Tesla V100 GPU.

\end{document}